\documentclass[aps,pra,twocolumn,groupedaddress,showpacs]{revtex4}

\usepackage{graphics}
\usepackage{amssymb}

\begin{document}

\title{Na-montmorillonite hydrates under ethane rich
reservoirs:\\
$NP_{zz}T$ and $\mu P_{zz}T$ simulations}

\author{G. Odriozola}
\email[]{godriozo@imp.mx}

\author{J. F. Aguilar}
\email[]{aguilarf@imp.mx}

\author{J. L\'{o}pez-Lemus}
\email[]{jllemus@imp.mx}

\affiliation{Programa de Ingenier\'{\i}a Molecular, Instituto
Mexicano del Petr\'{o}leo, L\'{a}zaro C\'{a}rdenas 152, 07730
M\'{e}xico, D. F., M\'{e}xico}

\date{\today}
\begin{abstract}
Na-montmorillonite hydrates in presence of ethane molecules are
studied by means of hybrid Monte Carlo simulations in the
$NP_{zz}T$ and $\mu P_{zz}T$ ensembles. The $NP_{zz}T$ ensemble
allows us to study the interlaminar distance as a function of
water and ethane content. These data show clear plateaus for lower
ethane contents and mainly for water contents consistent with the
formation of a single water layer. In addition, from this ensemble
the structure for some of these interlaminar compositions were
analyzed. For systems containing few ethane molecules and water
enough to complete a single layer, it was observed that ethane
mainly situates close to the interlayer midplane and adopts a
nearly parallel arrangement to the clay surface. On the other
hand, the $\mu P_{zz}T$ ensemble allows us to determine the
interlaminar distance and water-ethane content for any specific
reservoir. Here, some important findings are the following: the partial exchange
of water by ethane molecules that enhances for decreasing the
water vapor pressure; the obtention of a practically constant
interlaminar space distance as a function of the water vapor
pressure; the conservation of ion solvation shells; the
enhancement of the water-ethane exchange for burial conditions;
and finally, the incapability for a dehydrated clay mineral to
swell in a dry and rich ethane atmosphere.
\end{abstract}

\maketitle

\section{Introduction}
\label{intro}

Beyond the role that clay minerals play in several environmental
and industrial processes, one major concern is how their presence
affect petroleum migration and oil-recovery \cite{Newman,North}.
In these processes, petroleum, natural gas, and their associated
brines interact with clay minerals leading to retention, release,
and selective adsorption of water, ions, and certain organic
species, affecting the formation permeability. In general, the
amount and kind of the adsorbed species depend on their chemical
potential, which in turn establishes the swelling of a specific
clay, for a given pressure and temperature. Hence, changes in the
local reservoir composition or conditions, due to, for example, an
oil-recovery process, can cause a destabilization of the formation
clay minerals. This destabilization, together with a high
percentage of clay mineral in the basin, may lead to a formation
damage, {\it i.~e.} diminution of the permeability. Therefore,
knowledge of the clay behavior in presence of petroleum and
natural gas compounds and its associated brines at reservoir
conditions is mandatory for a better understanding of such
processes.

One possible way to study these systems is by means of computer
simulations. In the last few years, this tool has been widely used
for studying the swelling of montmorillonite hydrates for
different interlayer cations
\cite{Skipper95c,Skipper98,dePablo01b,Sposito00,Boek95a,Boek03,Skipper97,Sposito01,Marry02,Young00}.
There are, however, few theoretical works devoted to study the
montmorillonite behavior under basin conditions
\cite{deSiqueira97,dePablo04,Odriozola04}, and there are also few
theoretical papers dealing with the interaction of
montmorillonite hydrates with petroleum species
\cite{Park04,Titiloye00,Teppen97,Sposito99}. Hence, questions such
as do any oil or natural gas compounds enter into the interlaminar
space at reservoir conditions?, how many and what kind of
molecules enter?, do they expel water molecules in the process?,
and how does this affect the interlaminar space?, still do not have clear
answers.

For tackling some of these questions we performed hybrid Monte
Carlo (HMC) simulations to study the behavior of
Na-montmorillonite hydrates at equilibrium with different
ethane-water reservoirs. Ethane was chosen simply because it is an
abundant natural gas and live-oil compound, and it is expected to
enter into the interlaminar space. In addition, the Na$^+$ ion was
chosen since it is the most abundant cation found in
cation-clay-water systems. Simulations were carried out in the
$NP_{zz}T$ and $\mu P_{zz}T$ ensembles. This last one has the
advantage of allowing a direct measurement of the amount of
molecules and the interlaminar distance at equilibrium under the
established conditions. The employed water model is based on the
TIP4P-water \cite{Jorgensen79} and the Na$^+$-water-clay
interactions are given by Boek {\em et~al.} \cite{Boek95b}. The
ethane parameters were taken from Nath {\em et~al.} \cite{Nath98}.
Two condition sets were studied. These are ground level, P$=1$ atm
and T$=298$K, and P$=600$ atm and T$=394$K, which corresponds to a
burial depth of 4 km (assuming average gradients of 30 K/km and
150 atm/km).

The paper is organized as follows. In Sec.~\ref{methods}, we
briefly describe the models and the methodology employed for
performing the simulations. The results are shown in
Sec.~\ref{results}. Finally, Sec.~\ref{summary} summarizes the
main results and extracts some conclusions.

\section{Methodology}
\label{methods}
\subsection{The model}
The Wyoming type montmorillonite cell given by Skipper {\em
et~al.} \cite{Skipper95a} was replicated in such a way to form a
$4 \times 2$ layer. This layer has lateral dimensions of
$L_x=21.12$ $\hbox{\AA}$ and $L_y=18.28$ $\hbox{\AA}$, and
thickness of $6.56$ $\hbox{\AA}$. To obtain a typical Wyoming type
montmorillonite, isomorphous substitutions of trivalent Al atoms
of the octahedral sites by divalent Mg atoms, and tetravalent Si
by trivalent Al atoms were made. In this way, Clay I of the work by
Ch\'{a}vez-P\'{a}ez {\em et~al.} \cite{dePablo01} was
obtained, with unit cell formula
Na$_{0.75}$$n$H$_2$O(Si$_{7.75}$Al$_{0.25}$)(Al$_{3.5}$Mg$_{0.5}$)O$_{20}$(OH)$_4$.
Two layers were considered in the simulation box to avoid system
size effects \cite{dePablo01}. The rigid TIP4P model was used for
water molecules \cite{Jorgensen79} and a flexible model was
considered for ethane molecules, just as described in the work by Nath {\em
et~al.} \cite{Nath98}. For the initial configuration, water
and ethane molecules were randomly placed in the interlaminar
spaces, and sodium ions were distributed in the interlayer
midplanes. Note that six sodium ions per interlaminar space are
needed to keep the system electroneutral. Periodical boundary
conditions were applied in the three space directions.

The site-site intermolecular interactions are given by a
Coulombian contribution plus a Lennard-Jones potential,
\begin{equation}\label{pot}
U_{ij}\!=\!\sum_{a,b}\! \left[ \frac{q_aq_b}{r_{ab}}+ 4
\epsilon_{ab} \left[ \left( \frac{ \sigma_{ab}}{r_{ab}}
\right)^{12} - \left( \frac{ \sigma_{ab}}{r_{ab}} \right)^{6}
\right] \right]
\end{equation}
where subindices $i$ and $j$ are molecular indexes, and $a$ and
$b$ run over all sites of each molecule. Here, $q_{a}$ and $q_{b}$
are the charges of the corresponding sites, $\epsilon_{ab}$ and
$\sigma_{ab}$ are site-to-site specific Lennard-Jones parameters,
and $r_{ab}$ is the inter-site distance. The Lennard-Jones
parameters for single sites are given in Table \ref{parameters}.
The bond stretching potential for ethane molecules is
$U(r)$$=$$K_r(r-b_{eq})^2/2$, with $K_r$=$191.765$ kcal/(mol
$\hbox{\AA}$) and $b_{eq}$$=$$1.54$ $\hbox{\AA}$ \cite{Shyamal98}.
The site to site Lennard-Jones parameters are given by the
Lorentz-Berthelot rules
\begin{equation}
\sigma_{ab}=\frac{\sigma_a + \sigma_b}{2},
\end{equation}
\begin{equation}
\epsilon_{ab}=\sqrt{\epsilon_a \epsilon_b}
\end{equation}

\begin{table}
\caption{\label{parameters} Lennard-Jones parameters for
ethane-water-clay-Na$^+$ interactions.}
\begin{ruledtabular}
\begin{tabular}{ccc}
$\;\;\;\;$ Sites & $\epsilon \!$ (kcal/mol) & $\sigma \!$ ($\hbox{\AA}$) $\;\;\;\;$ \\
\hline $\;\;\;\;$ O & 0.155 & 3.154 $\;\;\;\;$ \\
$\;\;\;\;$ H & 0.000 & 0.000 $\;\;\;\;$ \\
$\;\;\;\;$ Na & 1.570 & 1.740 $\;\;\;\;$ \\
$\;\;\;\;$ Si & 3.153 & 1.840 $\;\;\;\;$ \\
$\;\;\;\;$ Al & 3.153 & 1.840 $\;\;\;\;$ \\
$\;\;\;\;$ Mg & 3.153 & 1.840 $\;\;\;\;$ \\
$\;\;\;\;$ CH$_3$ & 0.1990 & 3.825 $\;\;\;\;$ \\
\hline
\end{tabular}
\end{ruledtabular}
\end{table}

We should point out that the oxygen-oxygen parameters' values were
chosen to be exactly those of the TIP4P water model. These values
were slightly changed in references
\cite{dePablo04,Boek95b,dePablo01}. In addition, the sodium ion
parameters were obtained by a fitting procedure that leads to a
relatively good agreement between the pair energies obtained from
equations (7) and (8) of Bounds's work \cite{Bounds85} and
equation (\ref{pot}). These are, in fact, also similar to those
reported by Boek {\em et~al.} \cite{Boek95b}. The fitting
procedure was performed by employing the Levenberg-Marquardt
algorithm and considering several positions and orientations of
the water-ion pair. To check the accuracy of the fitted
expression, a HMC simulation of 216 water molecules and a Na$^+$
ion (and a Cl$^-$ ion to keep the system electroneutral) was
performed. The obtained Na$^+$-oxygen and Na$^+$-hydrogen radial
distribution functions, g(r), and coordination numbers, n(r), were
very similar to those reported by Bounds \cite{Bounds85}. Finally,
parameters for Si were taken from Marry {\em et~al.}
\cite{Levesque02}. Parameters for Al and Mg were assumed to be
equal to those of Si.

The Ewald summation formalism was implemented for handling the
electrostatic interactions. For this purpose, we set the
convergence factor to $5.6/L_{min}$, being $L_{min}$ the minimum
simulation box side. Five reciprocal lattice vectors for the
directions along the shortest sides and six for the direction
along the largest side were set \cite{Alejandre94}. For the
Lennard-Jones contribution a spherical cutoff of $L_{min}/2$ was
considered. In addition, this contribution was corrected by the
standard methods for homogeneous fluids \cite{Allen}.

\subsection{Simulations}
\label{simulation} A hybrid Monte Carlo scheme was employed for
the simulations. This technique has the advantage of allowing
global moves while keeping a high average acceptance probability.
A global move in configuration space consists in assigning
velocities and integrating the system over phase space using some
discretization scheme \cite{Mehlig92}. Velocities are randomly
assigned from a Gaussian distribution in correspondence with the
imposed temperature and in such a way that total momentum equals
zero for both interlaminar spaces. The discretization scheme we
employed is a reversible multiple time scale algorithm based on
the Trotter expansion of the Liouville propagator, which has the
advantage of allowing to split forces into short and long range
ones \cite{Tuckerman92}. Hence, a short time step is employed to
generate the motion using the rapid varying short range forces,
and a long time step is used only to correct the velocities by
using the slow varying long range forces. The short range forces
were chosen as the Lennard-Jones contribution plus the real part
of the electrostatic potential, and the long range forces were set
as the reciprocal space contribution of the electrostatic
potential. This allows reducing the expensive calls to the
reciprocal space contribution. For diminishing time correlations,
a new configuration is obtained after ten integration steps. Once
the new configuration is generated, it is accepted with a
probability
\begin{equation}
P=\min\{1,\exp(-\beta \Delta \mathcal{H})\}
\end{equation}
where $\Delta \mathcal{H}$ is the difference between the
Hamiltonians associated to the new and previous configuration and
$\beta$ is the inverse of the thermal energy. Note that increasing
the time step will result in a larger deviation from total energy
conservation and so, in a lower average acceptance probability. We
fixed the long time step to eight times the short time step, and we
set the short time step to obtain an average acceptance
probability of 0.7 \cite{Mehlig92}. For more details about the
hybrid Monte Carlo algorithm see Mehlig {\em et~al.}
\cite{Mehlig92}.

For sampling in the $NP_{zz}T$ ensemble, after a trial change of
particles' positions, a box change is attempted in such a way that
the stress normal to the surface of the clays, $P_{zz}$, is kept
constant. For this purpose, box fluctuations are allowed only in
the $z$-direction and the probability for accepting a new box
configuration is given by
\begin{equation}
P\!=\! \min\{ 1,\exp [ - \beta ( \Delta \mathcal{U}+P_{zz} \Delta
V \! - N \beta ^{-1} \! \ln (V_n/V_o)) ] \}
\end{equation}
Here, $\Delta \mathcal{U}$ is the change in the potential energy,
$\Delta V$ is the volume change, $N$ is the total number of
molecules, and $V_n$ and $V_o$ are the new and old box volumes,
respectively \cite{dePablo01}.

For sampling in an open ensemble, the possibility of insertions
and deletions of water and ethane molecules has to be considered.
Water insertions and deletions were performed by Rosenbluth
sampling as explained elsewhere \cite{Hensen01}. The only
difference is that we account for the real part of the
electrostatic contribution for the hydrogen trial conformations
(including the non massive TIP4P site contribution). This is a
necessity since the TIP4P hydrogen sites do not have a
dispersion-repulsion contribution, in contrast to the MCY water
model they employed. Hence, the real part of the electrostatic
contribution appears directly in the Rosenbluth factors and must
not be included in the exponential part of equations (5) and (6)
of the work by Hensen {\em et~al.} \cite{Hensen01}. Similarly,
insertions and deletions of ethane molecules were also performed
by the Rosenbluth method. Here, the slight difference is that
ethane molecules are flexible. In this case, the method is clearly
explained elsewhere \cite{Frenkel}.

Commonly, the open ensemble employed in simulations is the Grand
canonical, {\it i.~e.}~the $\mu VT$ ensemble. This was
successfully employed to determine the stable hydration states of
the water-ion-clay systems, {\it i.~e.}~the number of water
molecules and the interlaminar space at equilibrium with a given
reservoir \cite{dePablo01b,dePablo04,dePablo01,Hensen01,Hensen02}.
For this purpose, it is necessary to measure the pressure as a
function of the interlaminar distance (volume) to determine the
states that produce the reservoir's pressure. The problem is that
this implies a large number of simulation runs. Another
possibility would be sampling in the $\mu P_{zz}T$ ensemble. In
this way, a single simulation run is employed for obtaining the
system at equilibrium with the reservoir's conditions.

For finding an algorithm capable for sampling such ensemble, it is
necessary to pay attention on the probability density of finding
the system in a particular configuration. This is deduced from the
partition function, which in this case reads as
\begin{widetext}
\begin{equation}
\mathcal{Q}_{\mu_{1} \mu_{2} P_{z\!z}\!T} \! \propto \!\!\!\!\!
\sum_{N_1,N_2=0}^{\infty} \!\!\!\! \frac{ \exp [\beta (\mu_1 N_1
\! + \! \mu_2 N_2)]}{\Lambda^{3N}N_1!N_2!} \!\! \int \!\! dV \exp
[-\beta P_{z\!z}V]V^N \!\! \int \!\! d\mathbf{s}^N \exp[-\beta
\mathcal{U}(\mathbf{s\!}^N)]
\end{equation}
\end{widetext}
where $\Lambda$ is the thermal de Broglie wavelength, subindexes 1
and 2 refer to components 1 and 2 (water and ethane), and
$\mathbf{s}^N$ are scaled coordinates \cite{Frenkel}. Hence, the
corresponding probability density of finding the system in a
particular configuration is given by
\begin{equation}
\mathcal{N}_{\mu_1 \mu_2 P_{z\!z}\!T} \! \propto \! \frac{ V^N
\exp \{-\beta [\mathcal{U}(\mathbf{s}^N)\!-\!\mu_1 N_1\!-\!\mu_2
N_2\!+\!P_{z\!z}V]\}}{\Lambda^{3N}N_1!N_2!}
\end{equation}
and so, the algorithm must sample this distribution. For this
purpose, it is easy to show that particle movements, insertions
and deletions, and box changes must be done as in typical $NVT$,
$\mu VT$ and $NP_{zz}T$ sampling \cite{Frenkel}. In particular,
after trying a change of particles' positions, we perform several attempts of
inserting-deleting water and ethane molecules. This is done by
randomly calling the four possible trials in such a way that all
calls are equally probable. This is important in order to guaranty
the establishment of detail balance. Since accepting insertions or
deletions are rare, we repeat this step 10 times or until
accepting any insertion or deletion. In case of refusing the 10
insertion-deletion trials, we performed a box trial move. In this
way, the system rapidly evolves to an equilibrium state that, in
general, depends on the initial conditions.

\section{Results}
\label{results}

In order to organize the results this section is split into two
parts. These are {\it Sampling in the $NP_{zz}T$ ensemble} and
{\it Sampling in the $\mu P_{zz}T$ ensemble}. Each part presents
the results for the conditions of ground level and for 4 km of
burial depth. These are $T$$=$ 298 K, $P$$=$ 1 atm and $T$$=$ 394
K, $P$$=$ 600 atm, respectively.

\subsection{Sampling in the $NP_{zz}T$ ensemble}

This ensemble allows measuring the interlaminar distance as a
function of the imposed number of water and ethane molecules for a
given temperature and pressure. This means that water and ethane
content may or may not correspond to real systems.

\begin{figure}
\resizebox{0.45\textwidth}{!}{\includegraphics{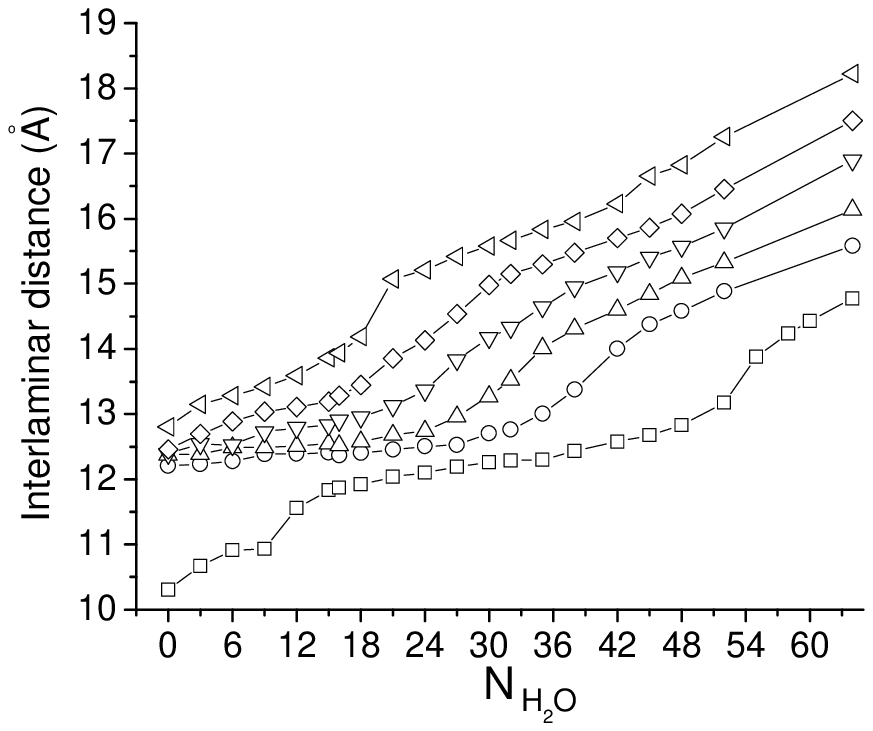}}
\caption{\label{dist_int_cfix} Interlaminar distance as a
function of the number of water molecules per interlaminar space.
Symbols {\tiny $\Box$}, $\circ$, {\tiny $\bigtriangleup$}, {\tiny
$\bigtriangledown$}, $\diamond$ and {\tiny $\lhd$} correspond to
0, 6, 9, 12, 15 and 18 fixed ethane molecules, respectively.}
\end{figure}

Fig.~\ref{dist_int_cfix} shows the interlaminar distance as a
function of the number of water molecules per clay sheet. Here,
the ethane content is fixed and the established conditions
correspond to ground level. Naturally, it is observed that as far
as the ethane content increases, the curves reach larger
interlaminar distances. Hence, the bottom curve corresponds to
zero ethane molecules and the uppermost curve to 18 ethane
molecules per interlaminar space. It should be pointed out that
this lowest curve is in good agreement with others reported
elsewhere \cite{dePablo01}, suggesting that the differences in
models and methods are not very important. This curve starts from
an interlaminar distance of 10.22 $\hbox{\AA}$, where ions are the
only specie. This dehydrated state is in good agreement with those
reported for the TIP4P water model
\cite{Hensen02,dePablo01,Boek95b} and slightly larger than the
reported experimental values (9.8 - 10.0 $\hbox{\AA}$)
\cite{Brindley}. The curve grows with increasing water content
reaching a plateau for 20-48 water molecules. This plateau yields
interlaminar spaces in the range 12.0-12.7 $\hbox{\AA}$, in good
agreement with experimental data. For larger water contents, a
second plateau is developed, which corresponds to a double water
layer.

When six ethane molecules are added to the interlaminar space, a
large plateau is developed in the water content range of 0-27,
which corresponds to interlaminar spaces ranging in 12.2-12.5
$\hbox{\AA}$. This means that as early as few ethane molecules
enter into the interlaminar space, they force it to yield
relatively large interlaminar distances and so, additional water
molecules easily enter without changing it. Furthermore, the fact
that only $27+6$ total molecules saturate the first layer implies
that ethane molecules occupy approximately the triple effective
volume than water. This is verified by the fact that for 9 and 12
ethane molecules the plateaus shorten, disappearing for 15 ethane
molecules per interlaminar space. Note that in each case the
plateaus are yielded for 0-18 and 0-6 water molecules,
respectively. We should also pointed out that in all cases the
plateaus produce interlaminar spaces in the range of 12.2-12.5
$\hbox{\AA}$. Hence, if ethane goes into the interlaminar space,
it is expected to displace water in the process without affecting
the interlaminar distance. In addition, these facts make us think
that for a single layer and in the most favorable situation, 12
ethane molecules may enter into the interlaminar space, displacing
approximately 36 water molecules in the process.

\begin{figure}
\resizebox{0.45\textwidth}{!}{\includegraphics{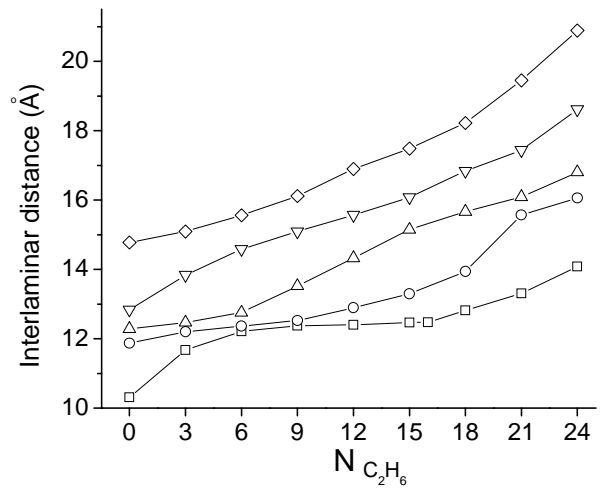}}
\caption{\label{dist_int_wfix} Interlaminar distance as a
function of the number of ethane molecules per interlaminar space.
Symbols {\tiny $\Box$}, $\circ$, {\tiny $\bigtriangleup$}, {\tiny
$\bigtriangledown$} and $\diamond$ correspond to 0, 16, 32, 48 and
62 fixed water molecules, respectively.}
\end{figure}

Similar conclusions can be drawn by studying the interlaminar
distance as a function of the number of ethane molecules keeping
fixed the number of water molecules. This was done and it is
presented in Fig.~\ref{dist_int_wfix}. As can be seen, the
plateaus for 0,16 and 32 water molecules correspond to the ranges
6-16, 3-9 and 0-3 ethane molecules, respectively. In addition, the
corresponding interlaminar distances are in the range 12.2-12.5
$\hbox{\AA}$. No other plateaus are seen for larger water
contents. In summary, these results confirm the conclusions
extracted from the preceding paragraph.

\begin{figure}
\resizebox{0.45\textwidth}{!}{\includegraphics{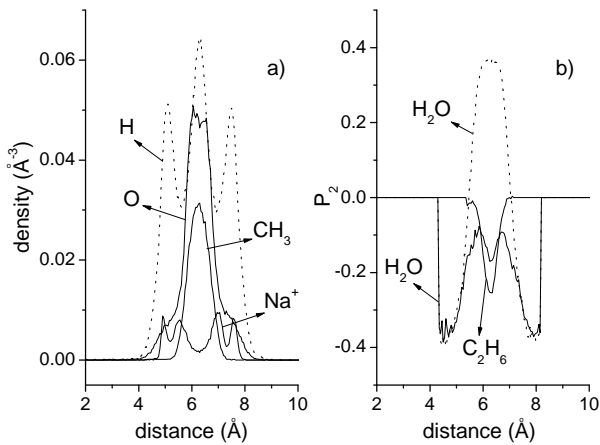}}
\caption{\label{perfil_w27_c6_km0} a) Density profiles of
oxygen, hydrogen, methyl, and sodium sites. They were obtained by
fixing 27 water and 6 ethane molecules. b) The corresponding
second Legendre polynomial order parameters for water and ethane
(see text).}
\end{figure}

Let us focus on the structure of the particular configuration of
27 water and 6 ethane molecules per interlaminar space. This is
shown in Fig.~\ref{perfil_w27_c6_km0}, where the density profiles
of oxygen, hydrogen, methyl, and sodium sites are plotted vs.~the
interlaminar distance from the clay surface. Note that oxygen,
hydrogen, and sodium profiles are nearly equal to those reported
by Ch\'{a}vez-P\'{a}ez {\em et~al.} \cite{dePablo01}. That is, we
observed the same peaks at the same distances for the different
species. The only difference is that the water peaks are shorter
due to their smaller content. Hence, the presence of ethane
molecules does not practically affect the structure of water and
ion sites. In other words, ethane molecules do not seem to interact
very much with water molecules and ions.

To study the orientation of ethane and water molecules as
a function of the interlaminar position, we compute the
statistical average of the second Legendre polynomial order
parameter
\begin{equation}
P_{2}=\langle 3\cos^{2}(\theta_i)/2-1/2 \rangle
\end{equation}
for all molecules, $i$. In case of water, the interlaminar oxygen
position and two angles were considered. The first one formed
between the dipole moment vector of the considered water molecule
and a vector normal to the clay layers (reference vector). The
other one is formed between the hydrogen-hydrogen direction and
the reference vector. These are shown in
Fig.~\ref{perfil_w27_c6_km0} b) as solid and dashed lines,
respectively. In case of ethane, the position of the center of
mass of each molecule and the angle formed between the
methyl-methyl direction and the reference vector were considered.
Hence, a positive value of $P_2$ means a tendency for the vectors
to get parallel to the normal of the clay sheets, whereas a
negative value of $P_2$ indicates a tendency for the vectors to
get perpendicular to the reference vector.

It is observed at the interlayer midplane that the water molecules
show a small tendency to have their dipole moment vectors normal
to the reference vector. Furthermore, they also show a tendency to
have their H-H-vectors parallel to the reference vector. For water
molecules closer to the clay surfaces, however, the H-H-vectors
turn perpendicular to the reference vector and their dipole
vectors become more normal to the reference vector. These results
are in good agreement with other works
\cite{Skipper95c,Odriozola04}. There are, however, some
differences of behavior between the TIP4P and the MCY-water
molecules previously studied by us \cite{Odriozola04}. That is,
for the TIP4P water we do not observe those few water molecules
that escape from the midplane and stick to the siloxane surface
that were observed for the MCY-water model. On the other hand,
ethane molecules were observed to show a tendency to align
parallel to the clay surfaces.

To study how the confinement and ethane presence affect the ions
solvation, the radial distribution functions and coordination
numbers for sodium-oxygen sites were built (not shown). Here,
total oxygen sites were split in water and clay oxygen sites so
that we can distinguish between the different contributions. It
was observed that all g(r) peak at 2.31 $\hbox{\AA}$, leading to a
total coordination number of 5.9 and a partial coordination number
for the water molecules of 3.7. Hence, sodium ions lose on average
more than two water molecules of their first shell to coordinate
with clay oxygen sites. In addition, the total coordination number
of 5.9 suggests that sodium ions are far from ethane molecules,
indicating that ethane molecules are being left aside. We should
also mention that since oxygen-oxygen separations on the clay
surface are significantly smaller than oxygen-oxygen separations
for waters of hydration shells around sodium ions, the total
coordination numbers may be somewhat inflated. This may question
the validity of some of these conclusions.

\begin{figure}
\resizebox{0.45\textwidth}{!}{\includegraphics{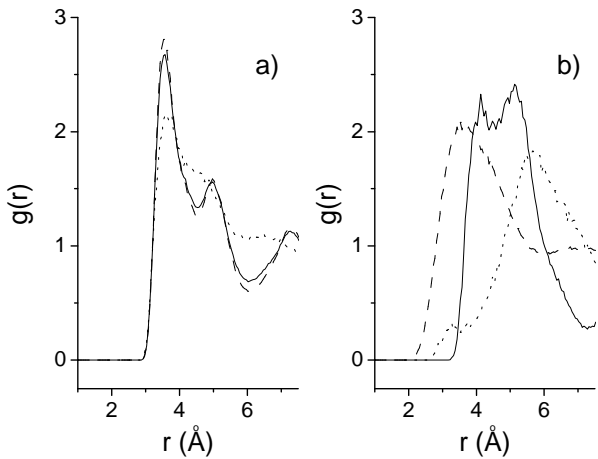}}
\caption{\label{gdr_C-O_C-C_C-H_C-I_lam} a) Radial
distribution functions for methyl-oxygen sites. Dotted, dashed and
solid lines correspond to water, clay and total oxygen sites,
respectively. b) Radial distribution function for methyl-sodium
(dotted line), methyl-hydrogen (dashed line) and methyl-methyl
(solid line) sites.}
\end{figure}

The radial distribution functions for the methyl-oxygen, -sodium,
-hydrogen and -methyl sites were also obtained and are shown in
Fig.~\ref{gdr_C-O_C-C_C-H_C-I_lam}. In case of oxygen sites, they
were again split into the corresponding water and clay
contributions. It is observed that all first methyl-oxygen peaks
are situated at 3.55 $\hbox{\AA}$ as well as the methyl-hydrogen
peak. Such structures are reminiscent of those observed in
hydrophobicity investigations and are a result of water-water
hydrogen bonding, since it is observed a lack of hydrogen atoms
pointing toward the nonpolar solute. Note that the second
methyl-oxygen peak appears at 4.97 $\hbox{\AA}$, corresponding to
the farthest methyl of the ethane molecule. This is also seen for
the methyl-methyl distribution function, where a double peak at
4.13 $\hbox{\AA}$ and 5.13 $\hbox{\AA}$ is observed. It should be
pointed out that the peak for the methyl-sodium distribution
function appears as far as 5.67 $\hbox{\AA}$. This means that, in
general, the water coordination shell that surrounds ions prevents
the ethane molecules to approach them. Nevertheless, a very small
shoulder appears at 3.25 $\hbox{\AA}$, meaning that a few ion
shells were partially broken or that the number of water molecules
were not enough to complete them. This may explain the 5.9
coordination number of sodium ions by oxygen atoms, instead of the
6.0 value we found for bulk.

\begin{figure}
\resizebox{0.45\textwidth}{!}{\includegraphics{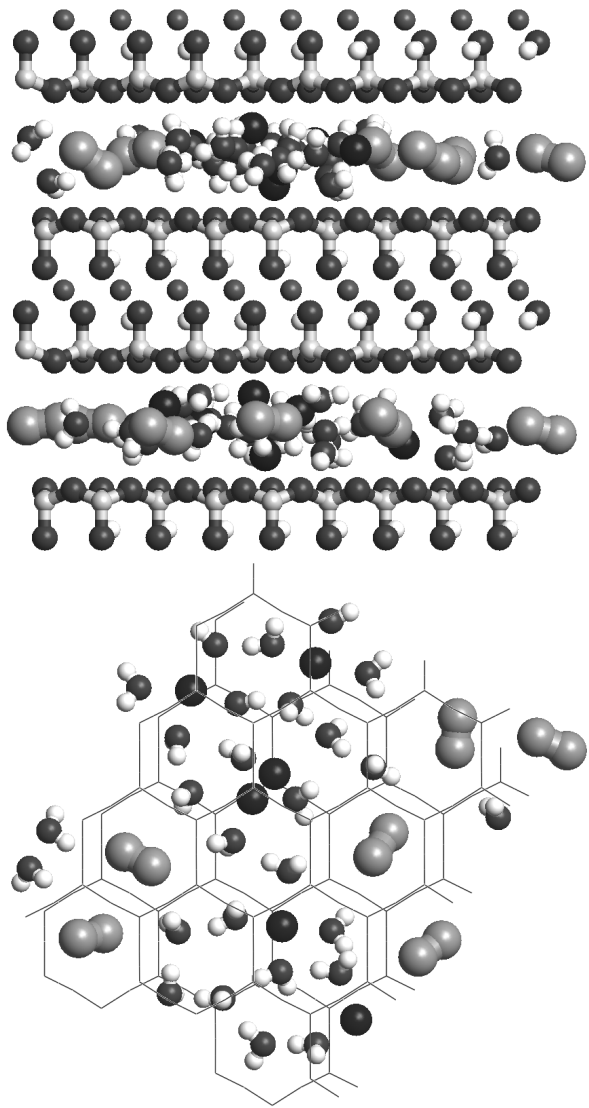}}
\caption{\label{snapshot} Snapshot of an equilibrated system
having 27 water molecules and 6 ethane molecules per interlaminar
space. Here, H are white, CH$_3$ are light gray, O are dark gray
and Na sites are black. Two views of the same configuration are
shown, a side view at the top and a top view at the bottom. Note
that for clarity, the bottom view shows only the topmost layer,
representing the clay structure by lines.}
\end{figure}

Fig.~\ref{snapshot} depicts some observations mentioned above,
where a snapshot of this system is presented. Here, a side view
shows how water molecules coordinate with ions in the middle of
the topmost layer displacing ethane molecules. The same topmost
layer configuration is also seen as a top view, where the
solvation of sodium ions and the isolation of the ethane molecules
are more evident. This also happens in the lowest layer, although
it is not clearly shown. Furthermore, it should be noted that
ethane molecules are located practically in the middle of the
interlaminar space where they are parallel to the clay surfaces,
as previously mentioned.

\begin{figure}
\resizebox{0.45\textwidth}{!}{\includegraphics{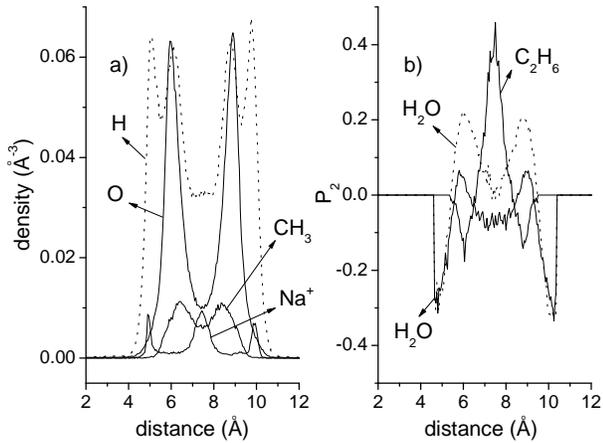}}
\caption{\label{perfil_w52_c6_km0} a) Density profiles of
oxygen, hydrogen, methyl, and sodium sites. They were obtained by
fixing 52 water and 6 ethane molecules. b) The corresponding
second Legendre polynomial order parameter for water and ethane.}
\end{figure}

Fig.~\ref{perfil_w52_c6_km0} shows the density profiles of the
interlaminar sites for the system containing 52 water and 6 ethane
molecules. It is clearly seen that two well defined oxygen peaks
indicate the formation of a double water layer. Hydrogen atoms
show two peaks for each oxygen peak, a larger one close to the
clay surface, and a smaller one almost overlapping its
corresponding oxygen peak. Sodium ions show three peaks, a larger
one at the interlayer midplane and two smaller ones close to the
clay surfaces. These observations are practically equal to those
found elsewhere \cite{dePablo04}. Again, the only difference seems
to be the height of the peaks, which is easily explained by the
smaller water concentration in the interlayer.

The orientation of water molecules is similar to the one layer
case. It is observed a dipole moment vector and a H-H-vector
mostly parallel to the clay surface for those water molecules
close to it. For the water molecules placed over the oxygen peak,
it is obtained a positive second Legendre order parameter for the
H-H-vector and even a slightly positive one for the dipole moment
vector. This means that both tend to align perpendicular to the
clay surface. On the other hand, the ethane molecules behave quite
differently than in the previous case. Here, although they still
situate in the midplane of the interlaminar space, one CH$_3$ site
situates closer to the upper clay surface and the other one
situates closer to the lower. Hence, and as can be seen in
Fig.~\ref{perfil_w52_c6_km0} b), ethane molecules tend to align
perpendicular to the clay surfaces. This explains the two CH$_3$
peaks shown in Fig.~\ref{perfil_w52_c6_km0} a).

\begin{figure}
\resizebox{0.45\textwidth}{!}{\includegraphics{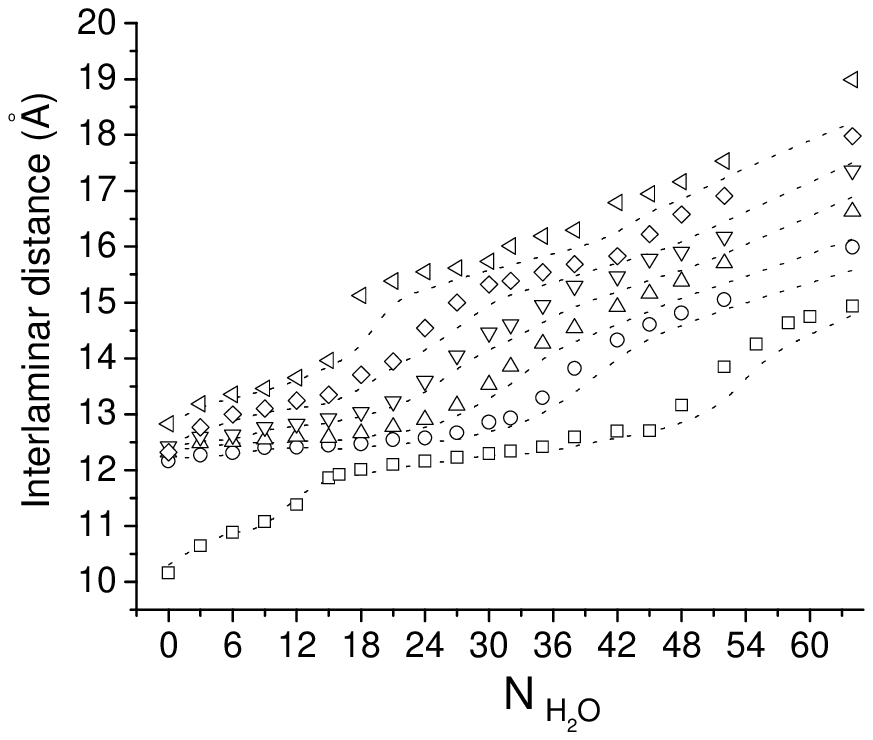}}
\caption{\label{dist_int_cfix_4km} Interlaminar distance as a
function of the number of water molecules per interlaminar space.
Symbols {\tiny $\Box$}, $\circ$, {\tiny $\bigtriangleup$}, {\tiny
$\bigtriangledown$}, $\diamond$ and {\tiny $\lhd$} correspond to
0, 6, 9, 12, 15 and 18 fixed ethane molecules, respectively. The
data is obtained for 4 km of burial depth. The corresponding data
obtained for ground level conditions (Fig.~\ref{dist_int_cfix})
are presented as dotted lines.}
\end{figure}

\begin{figure}
\resizebox{0.45\textwidth}{!}{\includegraphics{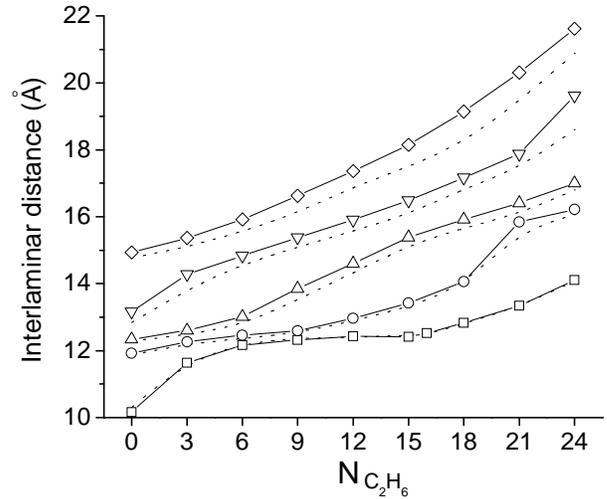}}
\caption{\label{dist_int_wfix_4km} Interlaminar distance as a
function of the number of ethane molecules per interlaminar space.
Symbols {\tiny $\Box$}, $\circ$, {\tiny $\bigtriangleup$}, {\tiny
$\bigtriangledown$} and $\diamond$ correspond to 0, 16, 32, 48 and
62 fixed water molecules, respectively. The data is obtained for 4
km of burial depth. The corresponding data obtained for ground
level conditions (Fig.~\ref{dist_int_wfix}) are presented as
dotted lines.}
\end{figure}

The interlaminar distance as a function of the number of water
molecules for 4 km of burial depth is shown in
Fig.~\ref{dist_int_cfix_4km}. The different data series
correspond, as in Fig.~\ref{dist_int_cfix}, to different ethane
contents. This figure also shows as dotted lines the data obtained
for ground level conditions. This is just to make easier the
comparison. As can be seen, the same trend is observed for both
conditions. The only difference is that for 4 km of burial depth
the interlaminar distances are larger than for the ground level
case. This difference is not very pronounced in any case, but
seems to be larger for those systems containing a larger number of
molecules. Similarly, Fig.~\ref{dist_int_wfix_4km} shows the
interlaminar distance as a function of the number of ethane
molecules, keeping the number of water molecules fixed.
Furthermore, and as in the previous figure, the data obtained for
ground level is presented as dotted lines. Again, it is observed
that the tendencies are similar although larger interlaminar
distances are obtained for the largest temperature and pressure
case. These findings are similar to those previously reported by
us \cite{Odriozola04}.

\begin{figure}
\resizebox{0.45\textwidth}{!}{\includegraphics{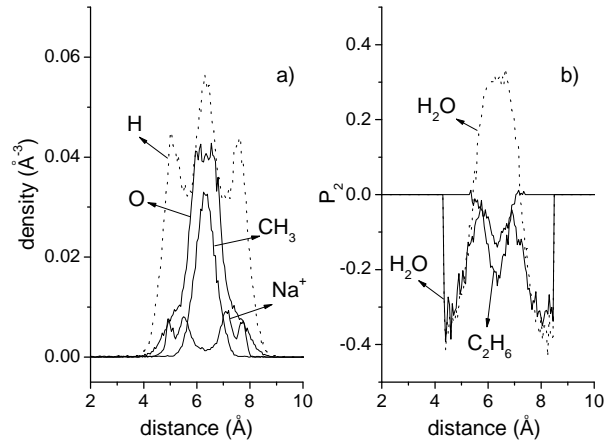}}
\caption{\label{perfil_w27_c6_km4} a) Density profiles of
oxygen, hydrogen, methyl, and sodium sites. They were obtained by
fixing 27 water and 6 ethane molecules, and for 4 km of burial
depth. b) The corresponding second Legendre polynomial order
parameter for water and ethane.}
\end{figure}

Finally, the density profiles and second Legendre polynomial order
parameters for the different sites are shown in
Fig.~\ref{perfil_w27_c6_km4}, for the system having 27 water and 6
ethane molecules and for 4 km of burial depth. As can be seen,
both, density profiles and order parameters are very similar to
those obtained for ground level conditions (see
Fig.~\ref{perfil_w27_c6_km0}). Nevertheless, the peaks are wider
and shorter than those obtained for ground level conditions,
indicating a larger disorder of the interlayer structure. This is
in good agrement with previous simulations
\cite{Odriozola04,dePablo04} and experiments \cite{Skipper00}.

\subsection{Sampling in the $\mu P_{zz}T$ ensemble}

As mentioned in section \ref{simulation}, this ensemble allows
obtaining the interlaminar distance, water content and ethane
content for the studied system at equilibrium with any given
reservoir's conditions. For that purpose, the chemical potential
of the species contained in the reservoir at given conditions must
be known. Hence, expression $\beta \mu$$=$$\beta
\mu_0$$+$$\ln{(p/p_0)}$ was used, where $p_0$ is the vapor
pressure at equilibrium with liquid water whose chemical potential
is $\mu_0$, and $p$ is the vapor pressure. For the TIP4P water
model and for T$=$298 K and P$=$1 atm, we obtained $\beta \mu_0$
$=$ -17.4 by simulating 200 water molecules in a cubic box. This
value is in good agreement with others reported for the same model
\cite{dePablo01}. For the studied burial conditions, {\it i.~e.},
for T$=$394 K and P$=$600 atm, we obtain $\beta \mu_0$ $=$ -13.4.

For obtaining the water chemical potential, we employed the
Rosenbluth insertion method as described in detail elsewhere
\cite{Hensen01,Frenkel}. For this purpose, $k_o$$=$$100$ oxygen
trial sites are randomly chosen inside the simulation box and the
quantity $\exp[- \beta u_{i}]$ is evaluated for each case. Here,
$u_{i}$ is the potential energy for a particular oxygen site that
comes only from the Lennard-Jones contribution. Immediately after,
a favorable oxygen site is selected by assigning to each site the
probability $P_{i}$$=$$\exp[- \beta u_{i}]/W_o$ with
$W_o$$=$$\sum_{i}\exp[- \beta u_{i}]$. For this selected oxygen
site, $k_h$$=$$20$ hydrogen configurations are tried while the
quantity $\exp[- \beta u_{j}]$ is calculated. In this case,
$u_{j}$ contains the contribution of the real part of the
electrostatic potential. Hence, both hydrogen atoms and the
charged non massive site of the TIP4P molecule contribute to
$u_{j}$. Again, a favorable orientation is selected by assigning
to each one the probability $P_{j}$$=$$\exp[- \beta u_{j}]/W_h$
where $W_h$$=$$\sum_{j}\exp[- \beta u_{j}]$. Once the TIP4P
molecule is fully inserted, the reciprocal space contribution to
the electrostatic potential, $u_r$, is evaluated. Finally, the
Rosenbluth weight factor is $W_{H_2O}$$=$$W_oW_h \exp[-\beta
u_r]/(k_ok_h)$. Since we perform the sampling from a $NP_{zz}T$
ensemble, the chemical potential finally reads \cite{Frenkel}
\begin{equation}
\beta \mu = - \ln \left ( \frac{1}{\beta P_{zz} \Lambda ^3} \right
) - \ln \left ( \frac{\beta P_{zz}}{N\!+\!1} \langle V W_{H_2O}
\rangle \right )
\end{equation}
where $\langle \cdots \rangle$ is the average obtained over 100000
configurations in which 10 water insertions per configuration were
done.

On the other hand, we use the relationship $\beta \mu$$=$$\beta
\mu_{id. gas}^0$$+$$\ln{( \beta p_e/ \Phi)}$ where $\mu_{id.
gas}^0$ is the ideal gas chemical potential of the reference
state, $p_e$ is the ethane partial pressure, and $\Phi$ is the
ethane fugacity coefficient \cite{Frenkel}. This coefficient is
0.992 for ground level conditions (taken from Din \cite{Din61})
and 0.536 for 4 km of burial depth (extrapolation of the data
given by Din \cite{Din61}).

\begin{figure}
\resizebox{0.45\textwidth}{!}{\includegraphics{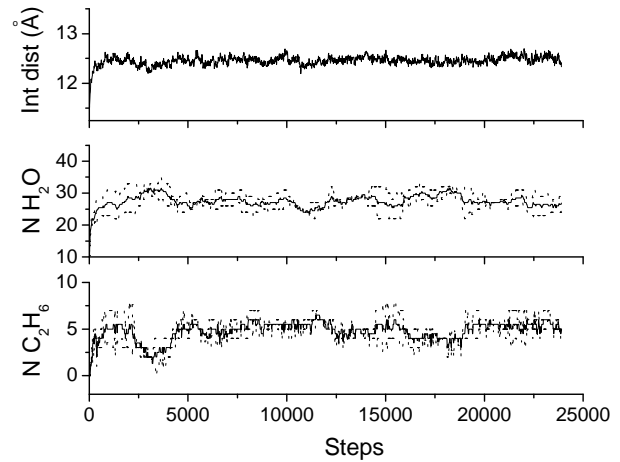}}
\caption{\label{run} Typical evolution of the interlaminar
space, water, and ethane content. Solid lines are averages whereas
dashed lines correspond to each interlaminar space. The
established conditions were T$=$298 K, P$=$1 atm, $p/p_0$$=$0.1
and $p_e=$0.893P. Initial conditions were established as 11.5
$\hbox{\AA}$ of interlaminar space and 10 water molecules.}
\end{figure}

In this way, for ground level conditions and for a reservoir
having $p/p_0=$ 0.1 and $p_e=$ 0.893P, Fig.~\ref{run} shows how
the system reaches equilibrium starting from a configuration of 10
water molecules and an interlaminar distance of 11.5 $\hbox{\AA}$.
For this case, it is observed that equilibrium is reached after a
few time steps, yielding an average interlaminar distance of 12.46
$\hbox{\AA}$, an average water content of 27.6 molecules, and an
average ethane content of 4.88 molecules per interlaminar space.
It should be mentioned that other runs need a larger number of
time steps for achieving equilibrium and so, runs must be
monitored in order to establish the step range to average. In this
case averages were performed in the range 5000-23000 steps. This
same procedure was then repeated systematically for obtaining
swelling curves for different reservoir and initial conditions.

\begin{figure}
\resizebox{0.45\textwidth}{!}{\includegraphics{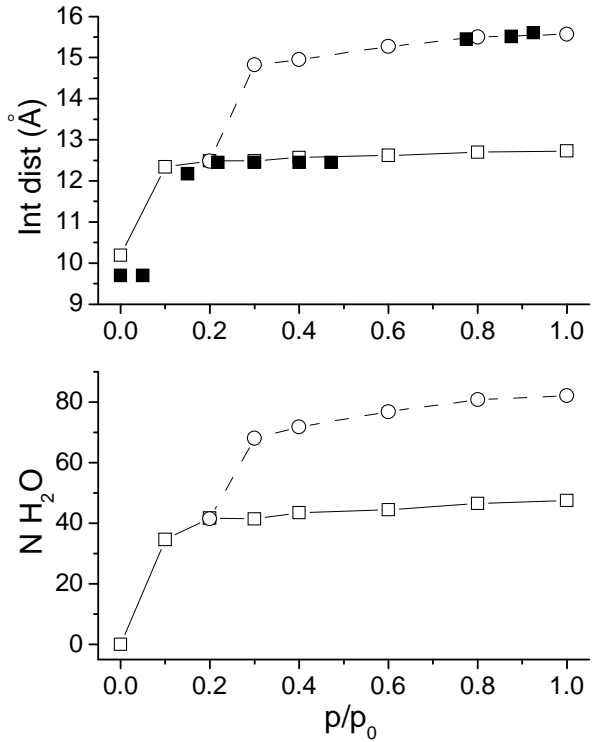}}
\caption{\label{id_n-vr-ph2o} Fig.~\ref{id_n-vr-ph2o}: Interlaminar distance and number of water molecules per interlaminar space as a function of
water vapor pressure and for ground level conditions. The ethane
partial pressure was set to zero. Symbols {\tiny $\Box$} and
$\circ$ correspond to the data obtained by simulations and
starting from 10 and 60 water molecules, respectively. Symbol
{\tiny $\blacksquare$} corresponds to experimental data reported
by Brindley and Brown \cite{Brindley}.}
\end{figure}

Fig.~\ref{id_n-vr-ph2o} was built for ground level conditions and
for a zero ethane partial pressure. It shows two swelling curves
which differ on the initial conditions. One starts from an
interlaminar distance of 11.5 $\hbox{\AA}$ and 10 randomly placed
water molecules. The other one starts from an interlaminar
distance of 16.0 $\hbox{\AA}$ and 60 water molecules. It can be
seen that for $p/p_0$ $\leq$ 0.2 both curves coincide, yielding
interlaminar distances and water contents lower than 12.5
$\hbox{\AA}$ and 42 molecules, respectively. In particular, for
$p/p_0$ $=$ 0, no water molecules and 10.22 $\hbox{\AA}$ of
interlaminar distance were obtained. For $p/p_0$ $\geq$ 2, the
curves separate producing two different plateaus. One corresponds
to the formation of a single water layer and the other one to the
formation of a double layer. This first plateau yields values of
the interlaminar distance in the range 12.34-12.73 $\hbox{\AA}$,
having 34.6-47.5 molecules of water content. The double layer
plateau yields 14.82-15.57 $\hbox{\AA}$ and 68.0-82.1 molecules,
respectively. The interlaminar distance values are in good
agreement with others reported by experiments and simulations
\cite{Hensen02,dePablo01,Boek95b,Brindley}. The water contents
agree with the ones reported for the TIP4P water model and
obtained by $\mu VT$ simulations \cite{dePablo01}. Nevertheless,
they are larger than others \cite{Hensen02}.

The presented swelling curves are similar to those reported by
Hensen and Smit \cite{Hensen02}. Both predict a single layer for
$p/p_0$ $\leq$ 0.2 and two layers for the range $p/p_0=$ 0.3-0.55.
For larger $p/p_0$, our data still predict the two configurations
to be stable and theirs show that the single layer is not stable
anymore. This may be the main difference between both results. On
the other hand, for a similar water-clay model and sampling from a
$\mu VT$ ensemble, Ch\'{a}vez-P\'{a}ez {\em et~al.}
\cite{dePablo01} also found that both configurations are stable,
in good agreement with our data. Finally, swelling experimental
data taken from Brindley and Brown's work \cite{Brindley} were
also included in the interlaminar distance graph, in order to be
compared with our results. As can be seen, the agreement is very
good.

\begin{figure}
\resizebox{0.45\textwidth}{!}{\includegraphics{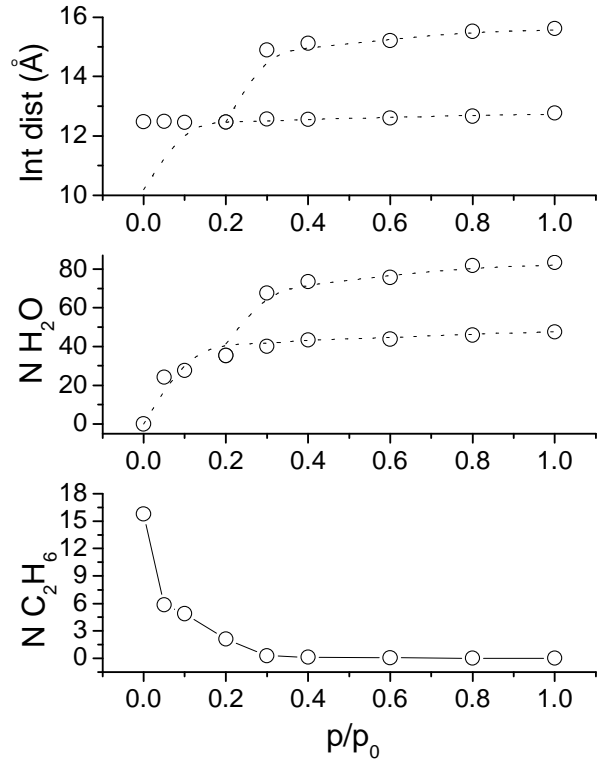}}
\caption{\label{id_n-vr-ph2o_pch} Interlaminar distance,
number of water molecules, and number of ethane molecules per
interlaminar space as a function of water vapor pressure. The
ethane partial pressure, $p_e$, was set to 0.893P and ground level
conditions were established. Dotted lines correspond to $p_e=$ 0,
which were included to make easy the comparison.}
\end{figure}

A similar study was made to determine the influence of a ethane
rich reservoir on the interlaminar space. For this purpose, a
partial ethane pressure of 0.893P was set for the same ground
level conditions. The obtained results are shown in
Fig.~\ref{id_n-vr-ph2o_pch} together with the curves already shown
in Fig.~\ref{id_n-vr-ph2o}. It is seen that, in general, they are
very similar, {\it i.~e.}~there is not a very pronounced
interaction between the ethane atmosphere in contact with the
clay-water system. In fact, for $p/p_0$ $\geq$ 0.4 the largest
ethane content observed was 0.12 molecules per interlaminar space.
This is equivalent to just one ethane per 350 water molecules.
Furthermore, for the double-layer cases the ethane content is even
lower, and this is why we did not include it in the plot. Moreover,
the ethane proportion decreases with increasing the vapor pressure
yielding zero for $p/p_0$ $\geq$ 0.8. Consequently, those points
obtained for $p/p_0$ $\geq$ 0.4 practically show the same
interlaminar distance and water content previously found.

Nevertheless, for $p/p_0$ $<$ 0.4 a clear
contribution of ethane to the interlaminar space can be seen. This
contribution turns more important by decreasing the vapor
pressure. This means that some water molecules are being replaced
by ethane molecules, keeping the interlaminar distance almost
constant. The exchange ratio is close to 3:1, as found in the
preceding section. For the extreme case of $p/p_0$$=$0, the whole
water layer is replaced by a layer of 16 nearly parallel ethane
molecules, which leads to a practically equal interlaminar
distance. It should be mentioned that, for this particular case,
the simulation was started from 12.5 $\hbox{\AA}$ of interlaminar
space and 10 ethane molecules, instead of 11.5 $\hbox{\AA}$ and 10
water molecules. If started from this last initial condition, the
system quickly loses its water content producing a stable
dehydrated state having zero ethane molecules. We also observed
that this stable dehydrated state is destabilized by any small but
enough vapor pressure. This indicates that the presence of a small
amount of water, which widens the interlaminar space and solvate
ions, aids the entrance of ethane molecules. This finding agrees
with experimental evidence reported by Barrer \cite{Barrer}.

\begin{figure}
\resizebox{0.45\textwidth}{!}{\includegraphics{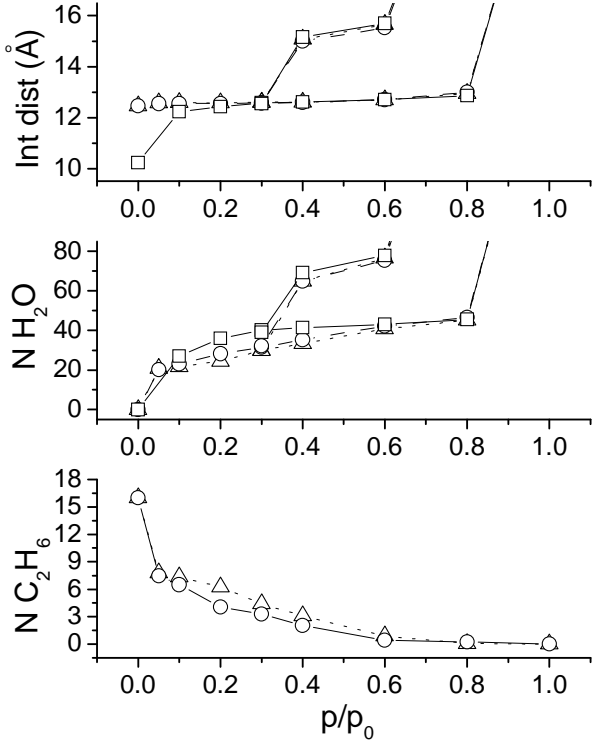}}
\caption{\label{id_n-vr-ph2o_pch_p} Interlaminar distance,
number of water molecules, and number of ethane molecules per
interlaminar space as a function of water vapor pressure and for 4
km of burial depth. Symbols {\tiny $\Box$}, $\circ$ and {\tiny
$\bigtriangleup$} correspond to $p_e$$=$ 0.0P, 0.107P and 0.268P,
respectively.}
\end{figure}

Similarly to Fig.~\ref{id_n-vr-ph2o_pch},
Fig.~\ref{id_n-vr-ph2o_pch_p} shows swelling curves but for 4 km
of burial depth. Here, three different ethane partial pressures
were considered. These are $p_e$$=$ 0.0P, 0.107P and 0.268P. For
$p_e$$=$ 0, some differences with respect to
the ground level case can be observed.

On the one hand, for $p/p_0$$=$1.0 neither a single layer nor a
double layer configuration are stable. For this particular case,
no matter what the established initial conditions, the interlaminar
space and water content monotonously increase as the simulation
evolves. They were ended after reaching 180 water molecules and,
therefore, we assume that a full hydration of the clay system was
obtained. A similar behavior was observed for $p/p_0$$=$0.8 and
for an initial configuration of 60 water molecules. When starting
this last system from an initial configuration of 10 water
molecules, however, the system reaches equilibrium yielding a
single water layer. This is in good agreement with the results
reported by Ch\'{a}vez-P\'{a}ez {\em et~al.} \cite{dePablo04},
where just a stable crowded single water layer configuration was
obtained for burial conditions. Nevertheless, they obtain this
behavior for $p/p_0$$=$1.0, instead of $p/p_0$$=$0.8. Their
employed conditions were 353 K and 625 bar, which correspond to
depths of 4.16 and 2 km, in terms of lithostatic and thermal
gradients, respectively. Hence, it seems that the difference of
temperature employed in both cases may explain the different
values of $p/p_0$ that produces a stable single layer and an
instable double layer.

On the other hand, it is observed that both curves predict the
same single water layer for $p/p_0$$\leq$0.3. Consequently, the
double layer configuration is stable just in the range $p/p_0=$
0.4-0.6. This result agrees with the experimental finding of a
dehydration of the interlaminar space from a double water layer to
a single one as the burial depth increases \cite{Skipper00}. In
addition, it also agrees with previous simulations of a similar
system but for the MCY water model, where the stability of the two
layer configuration was found to decrease with increasing burial
depth \cite{Odriozola04}.

Additionally, the interlaminar distances for single and double
layer configurations are similar to those corresponding to ground
level conditions. The water content, however, is slightly lower,
suggesting that water molecules occupy a larger effective volume.
This finding also agrees with the results reported for the MCY
water model \cite{Odriozola04}.

For $p_e$$=$ 0.107P and 0.268P, a similar effect to
the one found for ground level conditions is seen. That is, some water
molecules are replaced by ethane molecules, and this exchange is
enhanced for lower water vapor pressures. The extent of this
exchange, however, is much larger under burial conditions. This is
mainly explained by the higher ethane pressure, which leads to a
larger ethane activity. Naturally, the water-ethane exchange is
larger for $p_e$$=$ 0.268P than for 0.107P for all cases, as
expected. It is observed that even for $p/p_0=$0.6, 0.4 and 0.9
ethane molecules enter the interlaminar space for the single layer
case and for $p_e$$=$ 0.107P and 0.268P, respectively. This is
approximately equivalent to one ethane molecule per 100 and 45
water molecules. For the double layer cases, the water-ethane
exchange is not so pronounced. For $p_e$$=$0.268P and $p/p_0=$0.6,
only one ethane molecule per 220 water molecules were detected by
averaging a large number of $\mu P_{zz}T$ ensemble configurations.

In spite of the important exchange of water by ethane molecules
found for $p/p_0$$\leq$0.6, several water molecules still remain
in the system even for $p/p_0$ as low as 0.05. For this case and
for $p_e$$=$0.268P, an average of 20.8 water molecules were found
to be strongly attached to cations, impeding the entrance of more
than 7.8 ethane molecules. As for the ground level case, when
setting $p/p_0=$0, water is forced to leave the system and a single
layer of 16 ethane molecules is obtained. Again, this layer is not
obtained if the simulation is started from an initial
configuration close to the dehydrated state. This means that a
dehydrated clay containing just ions in its interlayer space and
in contact with a dry ethane rich reservoir does not allow the
entrance of ethane molecules.

\section{Conclusions}
\label{summary}

Hybrid Monte Carlo simulations in the $NP_{zz}T$ and $\mu P_{zz}T$
ensembles were used for studying the Na-montmorillonite hydrates
under ethane rich reservoirs. From the $NP_{zz}T$ study,
interlaminar distances as a function of the water and ethane
content were obtained. These curves show clear plateaus for lower
ethane contents and mainly for water contents consistent with the
formation of a single water layer.

On the other hand, from the $NP_{zz}T$ ensemble we also studied
the interlayer structure. Here it was observed, for systems
containing few ethane molecules and water enough to complete a
single layer, that ethane mainly situates close to the interlayer
midplane and adopts a nearly parallel arrangement to the clay
surface. Furthermore, it was observed that the water-ion-clay
interactions are practically unchanged by the presence of the
hydrocarbon, indicating that this component is being left aside.

In addition, the $\mu P_{zz}T$ ensemble allows us determining the
interlaminar distance and water-ethane content for any specific
reservoir's conditions. Hence, several swelling curves as a
function of water pressure were obtained. For ground level
conditions, they indicate that ethane molecules do not practically
enter the interlaminar space, unless one establishes a very low
water pressure. If this is the case, we also observed that water
molecules are flushed out by this entrance, leading to an almost
constant interlaminar space close to 12.5 $\hbox{\AA}$.

For the studied burial depth conditions, it was observed that the
exchange is enhanced, and that the ethane molecules are capable of
entering even for relatively large water pressures. On the other
hand, this exchange is not very pronounced for a double layer
configuration. Furthermore, it was observed that for very low
water pressures, the system holds water enough to keep the ions'
shells practically intact. Finally, our results also indicate that
a dehydrated and collapsed clay system is not capable to swell and
absorb ethane molecules if no water vapor is present in the
atmosphere.

\section{Acknowledgments}
This research was supported by Instituto Mexicano del Petr\'{o}leo
grant No D.00072. The authors would like to thank Dr.~L.~de Pablo for
his useful comments.

\end{document}